\documentclass[sigconf]{acmart}

\usepackage{graphicx}   
\usepackage{subcaption} 
\usepackage{soul}
\usepackage{flushend}

\AtBeginDocument{%
  }

\def\Section {\S}

\newcounter{todonumber}
\stepcounter{todonumber}

\newcounter{wnotenumber}
\stepcounter{wnotenumber}

\newcommand{\squishlist}{
 \begin{list}{$\bullet$}
  { \setlength{\itemsep}{0pt}
     \setlength{\parsep}{3pt}
     \setlength{\topsep}{3pt}
     \setlength{\partopsep}{0pt}
     \setlength{\leftmargin}{1.5em}
     \setlength{\labelwidth}{1em}
     \setlength{\labelsep}{0.5em} } }

\newcommand{\squishlisttwo}{
 \begin{list}{$\bullet$}
  { \setlength{\itemsep}{0pt}
     \setlength{\parsep}{0pt}
    \setlength{\topsep}{0pt}
    \setlength{\partopsep}{0pt}
    \setlength{\leftmargin}{2em}
    \setlength{\labelwidth}{1.5em}
    \setlength{\labelsep}{0.5em} } }

\newcommand{\squishend}{
  \end{list}}

\copyrightyear{2023}
\acmYear{2023}
\setcopyright{acmlicensed}\acmConference[HotCarbon '23]{2nd Workshop on
Sustainable Computer Systems}{July 9, 2023}{Boston, MA, USA}
\acmBooktitle{2nd Workshop on Sustainable Computer Systems (HotCarbon
'23), July 9, 2023, Boston, MA, USA}
\acmPrice{15.00}
\acmDOI{10.1145/3604930.3605710}
\acmISBN{979-8-4007-0242-6/23/07}



\begin{document}

\title{On the Promise and Pitfalls of Optimizing Embodied Carbon}

\author{Noman Bashir, David Irwin, Prashant Shenoy}
\affiliation{%
  \institution{University of Massachusetts Amherst}
  \streetaddress{}
  \city{}
  \country{}}

\renewcommand{\shortauthors}{Bashir et al.}

\begin{abstract}
To halt further climate change, computing, along with the rest of society, must reduce, and eventually eliminate, its carbon emissions. Recently, many researchers have focused on estimating and optimizing computing's \emph{embodied carbon}, i.e., from manufacturing computing infrastructure, in addition to its \emph{operational carbon}, i.e., from executing computations, primarily because the former is much larger than the latter but has received less research attention.  Focusing attention on embodied carbon is important because it can incentivize i) operators to increase their infrastructure's efficiency and lifetime and ii) downstream suppliers to reduce their own operational carbon, which represents upstream companies' embodied carbon. Yet, as we discuss, focusing attention on embodied carbon may also introduce harmful incentives, e.g., by significantly overstating real carbon reductions and complicating the incentives for directly optimizing operational carbon.   This position paper's purpose is to mitigate such harmful incentives by highlighting both the promise and potential pitfalls of optimizing embodied carbon.
\end{abstract}

\maketitle

\section{Introduction}
\label{sec:introduction}
To halt further climate change, computing, along with the rest of society, must rapidly reduce, and eventually eliminate, its carbon emissions by transitioning to lower carbon energy sources, such as solar, wind, nuclear, geothermal, and hydro.  Historically, this transition has been slow due to the higher costs and lower reliability of low-carbon energy compared to burning fossil fuels.  The simplest solution for this problem is to increase the relative cost of burning fossil fuels by placing a price on carbon, e.g., via a carbon tax or cap-and-trade system, to provide a direct financial incentive for businesses to adopt low-carbon energy.   Such an incentive would be configurable based on the magnitude of carbon's price.  Many governments have adopted carbon taxes and cap-and-trade systems~\cite{article1}. Of course, since carbon pricing raises energy costs, it can hurt legacy carbon-based energy businesses.  As a result, many countries including the U.S. are unlikely to ever introduce a direct carbon pricing policy, and instead are using more indirect means.  For example, the recent U.S. Inflation Reduction Act takes an indirect approach to financially incentivizing lower carbon energy by providing various tax subsidies for actions that promote its use~\cite{ira}. 

\begin{figure*}
    \centering
    \includegraphics[width=\textwidth]{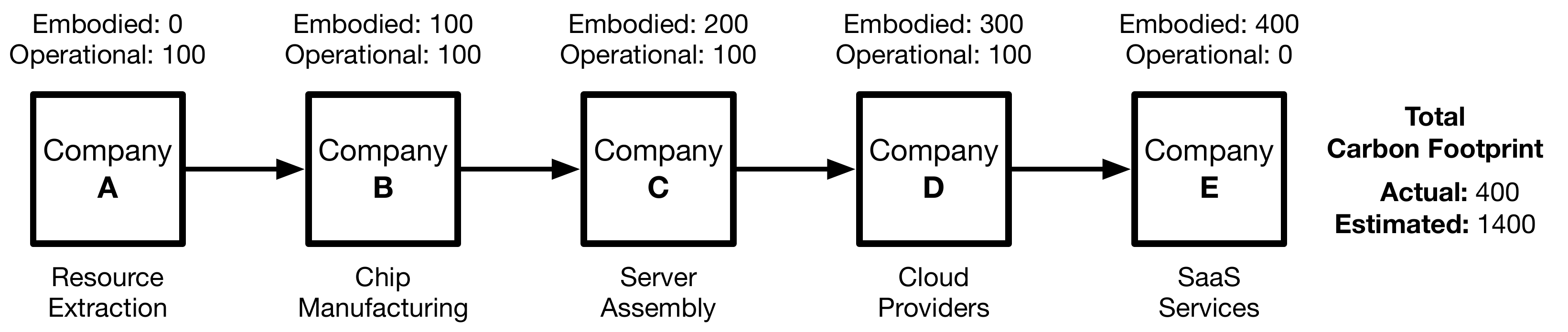}
    \vspace{-0.7cm}
    \caption{\emph{A toy supply chain with an actual carbon footprint of 400 units and an estimated carbon footprint of 1400 units.}}
    \vspace{-0.4cm}
    \label{fig:economy}
\end{figure*}
    
Since the financial incentives to adopt low-carbon energy introduced by the policies above are complex and likely not strong enough to reduce carbon emissions fast enough to avoid the worst outcomes of climate change, there has also been growing social pressure for companies to reduce their carbon footprint from their investors, customers, and employees, i.e., as part of Environmental, Social, and Governance (ESG) investing initiatives~\cite{esg}.  As a result, many companies now publicly report their annual estimated carbon emissions based on the Greenhouse Gas (GHG) protocol~\cite{ghg}, which is required in some countries and may soon be in the U.S.~\cite{article2}. The GHG protocol divides carbon and other emissions into Scopes 1, 2, and 3:  Scope 1 emissions derive from directly burning fuels and other chemicals, e.g., by company vehicles, generators, industrial processes, etc.; Scope 2 emissions derive from purchasing energy, e.g., from the electric grid; and Scope 3 emissions derive from all other aspects of a company's value chain, including carbon emissions from manufacturing the products and services a company uses.  Scopes 1 and 2 roughly represent a company's \emph{operational carbon}, while Scope 3 roughly represents its \emph{embodied carbon}.   Within computing, operational carbon is mostly Scope 2 and comes from using the grid to power IT equipment, while embodied carbon mostly comes from the manufacturing of computing infrastructure~\cite{chasing}.   

The increasing pressure for companies to report their carbon emissions has introduced a weak social incentive for them to reduce their operational and embodied carbon, which is influencing their behavior and optimizations.  These optimizations are complicated by the fact that the social incentive to reduce carbon emissions still often conflicts with their financial incentive to reduce cost.  Clearly, if there were a carbon tax, there would be no problem, as the social and financial incentives would align, and optimizing cost would also optimize carbon emissions. That is, companies would be financially incentivized to optimize their operational carbon, since lower carbon energy would incur lower carbon taxes and thus be cheaper; likewise, they would also be financially incentivized to optimize their embodied carbon, since products and services manufactured using lower carbon processes would be cheaper.  A carbon tax would also simplify accounting, as it often applies directly to the sale of fuel and other chemicals --- at a single point in the supply chain --- based on their carbon content.   Thus, when applying a carbon tax, there is less need to track carbon up and down the supply chain or even distinguish between embodied and operational carbon.   

Unfortunately, as mentioned above, most countries are unlikely to introduce a simple, direct, and meaningful carbon tax.  Given that, understanding how the incentives above will influence behavior and carbon optimizations is important.  In particular, there has been a recent focus in the computing research community on estimating and optimizing computing's embodied carbon, in part, because it is significantly larger than operational carbon but has received much less research attention~\cite{chasing,embodied1,embodied2,embodied3}.  Optimizing embodied carbon is important because it motivates operators to choose suppliers with a lower carbon footprint, as well as increase their own infrastructure's efficiency and lifetime to better amortize its embodied carbon. Such optimizations can influence the entire supply chain by indirectly incentivizing every downstream supplier to also reduce their carbon footprint -- both operational and embodied -- as well as increase their product lifetime.  By comparison, while optimizing operational carbon is more direct, as it reduces a single operator's emissions, it does not affect incentives up and down supply chain.  

The GHG protocol explicitly includes embodied emissions (roughly Scope 3) to ensure an holistic accounting of each company's carbon footprint.  As a result, even companies that achieve a low operational carbon still have some social incentive to lower their embodied carbon, e.g., by selecting manufacturers with lower operational carbon, increasing their infrastructure's lifetime, using less infrastructure by improving efficiency, etc. This essentially captures the purpose and promise of optimizing embodied carbon: similar to a carbon tax, it can provide an incentive (albeit a much weaker social one) for optimizations throughout the economy.   In addition, accounting for embodied carbon prevents companies from ``hiding'' their emissions by purposefully shifting them from operational to embodied.  As one example, a company might intentionally outsource its transportation needs to a third-party company, so that their associated emissions are not part of their operational carbon.  

Importantly, though, there are also a number of potential pitfalls when optimizing embodied carbon, especially in relation to operational carbon, that can introduce harmful incentives, which are not well understood by the research community. This position paper's purpose is to mitigate such harmful incentives by highlighting these pitfalls, summarized below, so future research can avoid them.  

\squishlist

\item {\bf Compounding Inaccuracy (\Section\ref{sec:inaccuracy})}.  Embodied carbon estimates are strictly less accurate than operational carbon estimates due to compounding inaccuracies across the supply chain. 

\item {\bf Double-counting Carbon (\Section\ref{sec:double})}.  Since one company's operational carbon is another company's embodied carbon, the same carbon emissions are counted multiple times, which leads to vastly overstating the total amount of embodied carbon. 

\item {\bf Embodied $>$ Operational (\Section\ref{sec:importance})}.  Nearly all companies in a complex economy with large supply chains will have high embodied-to-operational carbon ratios. However, this does not mean that embodied carbon is more important than operational carbon. 

\item {\bf Combining Metrics (\Section\ref{sec:combining})}.  Metrics that combine embodied and operational carbon are not meaningful, since, due to double-counting, embodied and operational carbon are measured on widely different scales and are thus incompatible. 

\squishend

Notably, the pitfalls above are general, and are broadly applicable across any industry.  However, the computing research community is particularly susceptible to these pitfalls due to its focus on automated software-driven optimization of embodied and operational carbon. We elaborate on the pitfalls above within the context of a toy supply chain and its operational and embodied carbon (\Section\ref{sec:economy}). We then discuss the implications for research moving forward (\Section\ref{sec:discussion}).  In doing so, we emphasize that optimizing embodied carbon is a critical component of carbon accounting (especially in the absence of a strong carbon tax), as the incentives it introduces are distinct from those introduced from optimizing operational carbon. 

\section{A Toy Supply Chain}
\label{sec:economy}

Figure~\ref{fig:economy} illustrates a toy supply chain for computing infrastructure.   At the root is a company A that extracts resources from the Earth, e.g., the various minerals required in production, and sells those resources to a company B that refines them and uses them to manufacture chips and other hardware.  Company B sells those chips and hardware to company C, which packages them together and assembles server platforms.  Company C in turn sells those servers to company D, a cloud provider that rents out remote access to them via virtual machines (VMs).  Finally, company E rents cloud VMs from company D to host a software-as-a-service platform, e.g., for video conferencing, messaging, file storage, etc.

For illustrative purposes, we assume our toy supply chain operates in isolation with a clear beginning and end, and that the supply chain above represents the entire economy.  That is, company A at the root requires no external inputs from other companies, and company E only sells its product to end-users and not other companies.  Of course, real supply chains do not have such clear beginnings and ends, as every company generally requires some inputs from others.  For example, in reality, extracting raw materials requires sophisticated equipment and fuel for excavation and transportation.  In practice, supply chains are more of a complex mesh of inter-dependent connections than a linear chain.  Modern supply chains are also broad and deep with most companies relying on numerous other companies to deliver goods and services to market.  

In general, the more complex a product, the more complex its supply chain.  Notably, technology products represent some of the modern economy's most complex products, and thus unsurprisingly their production necessitates massive and complex supply chains. For example, the Apple iPhone is made of hundreds of parts, e.g., the screen's glass, camera, chips, sensors, etc., from 191 suppliers across 43 countries and six continents~\cite{article3,article4}.  The supply chain for Tesla's electric vehicles is even larger with over 300 suppliers providing over 2000 parts~\cite{tesla1}.  Of course, the suppliers above also require external inputs from their own downstream suppliers, and thus only represent one level of depth in the iPhone and Tesla supply chains.  Given their size and complexity, companies generally do not have complete visibility into their full upstream and downstream supply chains. Supply chains are broad, in part, for robustness, as companies will often source components from multiple suppliers in case any one goes offline. Even so, modern supply chains are still fragile due to their size and the large physical distances between suppliers. This fragility became evident during COVID-19, as many supply chains broke down due to work and travel restrictions. 

In our toy supply chain, we assume each company incurs both operational and embodied carbon. Specifically, we assume each company consumes 100kg$\cdot$CO$_2e$ (or carbon equivalent), i.e., roughly Scopes 1 and 2, to produce their product as part of their operations.  We also assume company A has zero embodied carbon, since it is at the beginning of our toy supply chain and thus requires no external inputs.  As a result, company B incurs 100kg$\cdot$CO$_2e$ of embodied carbon, since it uses products from company A and must account for its carbon footprint.   Likewise, company C incurs 200kg$\cdot$CO$_2e$ embodied carbon, since it uses company B's products, while company's D and E incur 300kg$\cdot$CO$_2e$ and 400kg$\cdot$CO$_2e$, respectively, due to using products from their downstream suppliers.   

In general, the embodied carbon of any company $i$ represents the total carbon footprint (embodied and operational) of suppliers that are immediately downstream, i.e., based on the products purchased from that supplier.  That is, the carbon embodied in a product or service is the operational and embodied carbon emissions that were required to produce that product or service. Equivalently, the embodied carbon is also the sum of the operational carbon of all downstream suppliers.   In our example, since each company purchases 100\% of the products from the downstream company, its embodied carbon is simply the sum of its downstream supplier's own operational and embodied carbon. In practice, of course, a company's embodied carbon would need to be normalized relative to the amount and type of products it purchases. 

\section{Pitfall \#1: Compounding Inaccuracy}
\label{sec:inaccuracy}

Accurately and verifiably measuring operational and embodied carbon emissions, and then tracking them through the economy, is challenging.   For operational carbon, carbon information services, such as electricityMap~\cite{electricity-map} and WattTime~\cite{watttime}, have only recently provided some limited visibility into electricity's real-time carbon emissions.  However, their reported carbon emissions are unverifiable estimates based on models, not sensors, that infer carbon emissions from publicly-available data on each region's power plant characteristics, such as fuel type, and their real-time energy output.  These estimates are only available in regions that release data publicly, and also do not include other chemical processes in industry and agriculture that generate $\sim$8\% of U.S. emissions~\cite{eia1}. In short, while visibility into operational emissions is improving, it is far from complete, due to unknown model accuracy and a lack of transparency into emissions not derived from burning fossil fuels. 

Since embodied carbon is ultimately just a sum over operational carbon across the supply chain, accurately and verifiably measuring it suffers from the same issues as operational carbon above.  However, the problems above become compounded when measuring embodied carbon for companies higher in the supply chain, such as technology companies, since the accuracy relies on the accuracy of every downstream supplier.  That is, the inaccuracy increases as we move up the supply chain.  To illustrate, assume that every company in our toy example reports an operational carbon of 90kg$\cdot$CO$_2e$, or 10kg$\cdot$CO$_2e$ less than their actual operational carbon. In this case, company E at the top of the supply chain would report 360kg$\cdot$CO$_2e$ of embodied carbon, or 40kg$\cdot$CO$_2e$ less than their actual.  Since absolute errors accumulate up the supply chain, the scale of the error is a function of the supply chain's size, such that larger supply chains result in more error.   In general, there is no way to verifiably track the carbon provenance of products and services.   In addition, as mentioned in \Section\ref{sec:economy}, since companies generally do not have full visibility into their complete supply chain, they are not in a position to audit embodied carbon measurements. 
 
\noindent {\bf Key Takeaway:} \emph{Accurately and verifiably estimating embodied carbon is strictly more challenging than operational carbon due to the potential for compounding inaccuracies across the supply chain.}

\section{Pitfall \#2: Double-counting Carbon}
\label{sec:double}

Assuming our carbon accounting is accurate, the total carbon emissions in our toy economy above are 400 kg$\cdot$CO$_2e$, as it is simply the sum of each company $i$'s operational carbon ($C_o$).  However, notably, the total carbon footprint of company E alone, i.e., embodied plus operational carbon, is also 400 kg$\cdot$CO$_2e$, since its embodied carbon includes the carbon emissions from the entire supply chain.   Further, the sum of the total carbon footprint (operational and embodied) across all the companies is 1400 kg$\cdot$CO$_2e$, or $3.5$$\times$ the entire economy's carbon footprint. Obviously, this occurs because each company's embodied carbon is the sum of the operational carbon of every downstream supplier, such that the operational carbon of companies lower in the supply chain is counted multiple times. This example shows that holistic carbon accounting including embodied and operational carbon introduces a significant carbon multiplier effect.   That is, the same carbon emissions are not only double-counted, but are essentially multiplied by the length of the upstream supply chain.  While such double-counting of embodied carbon is acknowledged in the GHG protocol~\cite{ghg1,ghg2} and by researchers~\cite{article-scope3}, it is not widely appreciated.  

One significant problem with this multiplier effect is that it can lead to substantially overstating real carbon reductions.  For example, if company A reduces their operational carbon by 50\% (or 50kg$\cdot$CO$_2e$) due to increasing their use of renewable energy, it reduces the embodied carbon of every upstream company, such that each company can claim a 50kg$\cdot$CO$_2e$ reduction in their total carbon footprint (for a total carbon reduction of 250kg$\cdot$CO$_2e$ across the five companies).  Thus, any reductions in embodied carbon should be discounted, since multiple companies can claim ``credit'' for the same reduction. However, appropriately assigning credit for a reduction in embodied carbon can be difficult.  In the case above, likely only company A should receive the credit.  However, in another case, company D might improve its operational efficiency, leading it purchase fewer servers from company C and thus incur less embodied carbon.  In this case, company D should likely receive the credit for the reduction in embodied carbon. 

\begin{figure*}[t]
    \centering
    \includegraphics[width=\textwidth]{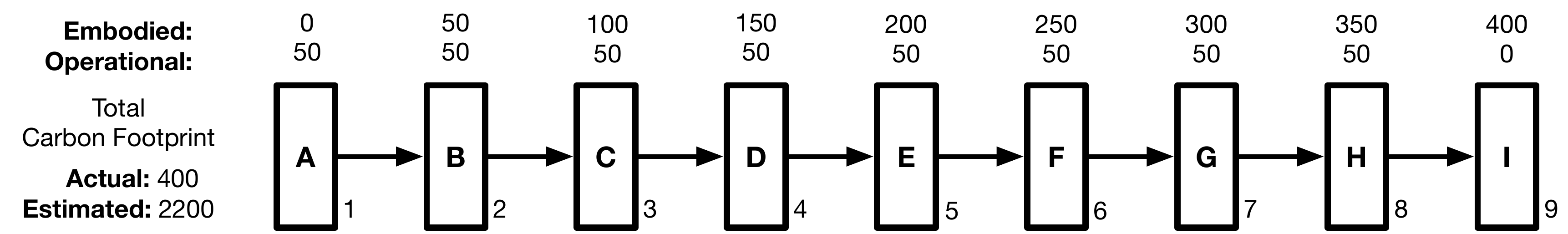}
    \vspace{-0.7cm}
    \caption{\emph{A longer supply chain with the same carbon footprint of 400 units, but total carbon footprint of 2200 units.}}
    \vspace{-0.4cm}
    \label{fig:example2}
\end{figure*}

The size of the multiplier effect is also a function of the supply chain's size. Thus, by restructuring the supply chain, we can increase the total embodied carbon without changing the operational carbon. For example, in Figure~\ref{fig:example2}, we extend our toy supply chain by dividing the work across 9 companies instead of 5, such that the first 8 companies in the supply chain consumed 50kg$\cdot$CO$_2e$ each.  In this case, the aggregate amount of embodied carbon would increase by 80\%, from 1000 kg$\cdot$CO$_2e$ to 1800 kg$\cdot$CO$_2e$.  Similarly, the total carbon footprint across all the companies in the supply chain would increase from 400 kg$\cdot$CO$_2e$ to 2200 kg$\cdot$CO$_2e$, or by 5.5$\times$.  As a result, even if total carbon emissions remain the same, the scale of embodied carbon can change based on the supply chain's characteristics and its division of labor.  Since, as mentioned above, the supply chain for technology products is especially large, i.e., hundreds of companies, it would suffer from a significant multiplier effect. This is harmful because it can provide the false impression that a company's reductions in real carbon emissions are much larger than they actually are, and may reduce the incentive for companies to make reductions in their own operational carbon.  That is, companies have a strong incentive to give the appearance, i.e., to investors, employees, regulators, and customers, of the largest possible reductions in carbon for the lowest possible cost.  Due to the multiplier effect above, reducing their embodied carbon is generally cheaper than reducing their operational carbon, since it requires much less reductions in others' operational carbon. 

In addition to the multiplier effect, the complexity of accounting for embodied carbon across the supply chain can also lead to incomplete analyses that may result in non-optimal decisions.  For example, a common mistake when analyzing the embodied and operational carbon of various combinations of renewables, storage, and grid energy is to account for the embodied carbon of renewables and storage, but not grid energy's embodied carbon, i.e., from building power plants, which makes renewables and storage appear comparatively worse for the environment.  In general, a full and accurate accounting of embodied carbon across the real economy -- with broad and deep supply chains -- remains challenging. 

\noindent {\bf Key Takeaway:} \emph{In aggregate, reductions in embodied carbon across the supply chain do not reflect real carbon reductions, since multiple companies are able to claim the same carbon reduction.}

\section{Pitfall \#3: Embodied $>$ Operational}
\label{sec:importance}

Another consequence of the carbon multiplier effect is that the total amount of aggregate embodied carbon claimed by the companies in our toy supply chain is much (2.5$\times$) larger than the total carbon emissions.  Clearly, the higher a company is in the supply chain, the larger its ratio of embodied-to-operational carbon. In general, the size of the ratio of embodied-to-operational carbon is mostly just a reflection of a company's position in the supply chain and the complexity of its product.  Since, as mentioned earlier, supply chains in the modern economy are less linear chains, and more of an interconnected mesh, essentially every company in a modern economy is ``high'' in its own supply chain, as nearly all companies depend on the products and services of many other companies.  Thus, it is not surprising that prior work has observed that embodied carbon for technology companies, which are especially high in the supply chain, is much larger than their operational carbon~\cite{chasing,embodied1,embodied2,embodied3}.  

Importantly, though, this observation that a company has a high embodied-to-operational carbon ratio \emph{does not mean} that reducing embodied carbon is more important than reducing operational carbon.  Indeed, embodied carbon \emph{is} operational carbon from another perspective.   Since one company's embodied carbon is just another company's operational carbon, this is akin to each company saying that reducing their operational carbon is less important than other companies in their supply chain reducing their operational carbon.    Again, this is harmful because it encourages companies to deflect responsibility and reduces the incentive for them to make potentially difficult and costly changes to reduce their own operational carbon. Due to the multiplier effect, companies essentially get much more ``bang for their buck'' when reducing embodied carbon.  Yet, reductions in operational carbon are arguably more meaningful, since they reflect real carbon reductions without a multiplier.

\noindent {\bf Key Takeaway:} \emph{In a complex economy with long inter-connected supply chains, most companies will have a high embodied-to-operational carbon ratio.  However, this does not mean that reducing embodied carbon is more important than reducing operational carbon.}

\section{Pitfall \#4: Combining Metrics}
\label{sec:combining}

When considering embodied carbon optimizations, a common approach is to introduce and optimize new metrics that combine embodied with operational carbon, as there are often tradeoffs between them.    For example, since renewable generation is highly variable, there is a tradeoff between embodied and operational carbon when provisioning it.  That is, provisioning more renewables and batteries to supplement grid energy reduces operational carbon, but with ever diminishing returns, while also increasing embodied carbon (from manufacturing the panels/batteries).  Similarly, provisioning more computing infrastructure enables systems to better exploit periods when low-carbon energy is available, and thus reduces operational carbon, but also increases their embodied carbon (from manufacturing servers).  Given the tradeoffs above, a simple combined optimization metric is the sum of embodied and operational carbon, i.e., the total carbon footprint.  Indeed, recent work has used such combined metrics~\cite{embodied2, junkyard-computing}, and a recent proposal for a Software Carbon Intensity (SCI) metric is essentially the average of software's embodied and operational carbon~\cite{sci}.

However, as mentioned above, due to the carbon multiplier effect, embodied and operational carbon are measured on very different scales, and thus are not really comparable. Recall that, in our toy supply chain, the total embodied carbon across the five companies is 1000kg$\cdot$CO$_2e$, while the total operational carbon (and also total carbon emissions) are only 400 kg$\cdot$CO$_2e$.   Thus, optimizing a combined embodied/operational metric per company will be biased towards reducing embodied carbon since it is much larger (due to our economy's complexity), even though, as mentioned above, those reductions are less meaningful than operational carbon reductions.  Further, the longer the supply chain, the larger the embodied energy, and thus the larger the bias.  Given the relationship between operational carbon, embodied carbon, and economic complexity, it is difficult to understand the macro effects of optimizing even a simple metric that additively combines operational and embodied carbon, much less more complex combinations~\cite{junkyard-computing,metrics}. In general, due to their different scales, embodied and operational carbon should be viewed as mathematically incompatible. 

\noindent {\bf Key Takeaway:} \emph{Since operational and embodied carbon are measured on different scales, they are incompatible. Metrics that combine embodied and operational carbon will be biased towards reducing embodied carbon due to the carbon multiplier effect.}

\section{Discussion}
\label{sec:discussion}

Embodied carbon is an important metric, as discussed in \Section\ref{sec:introduction}, that is a necessary part of holistic carbon accounting frameworks, since it provides an incentive for companies to reduce operational carbon throughout the supply chain and prevents them from ``hiding'' their operational emissions using third-party companies.   The pitfalls we describe are not intended to decrease the importance of embodied carbon, but to better contextualize it for those in the computing research community.   Indeed, embodied carbon is the same as operational carbon, but viewed from a different perspective. The key difference between embodied and operational carbon is that they introduce different incentives for reducing carbon emissions. 

Reducing operational and embodied carbon are both independently important, and thus are both an important focus of research.  For example, research on increasing the lifetime of equipment is an important direction for reducing the annualized embodied carbon of an organization. Similarly, organizations and individuals can also reduce their embodied carbon by buying upstream products with a lower embodied carbon, i.e., by substituting their current choices with  purchases of equivalent but lower-carbon options. Exposing data on products' embodied carbon will be important in not only enabling each company in a supply chain to source their inputs from greener upstream entities but also incentivizes them to reduce their own operational and embodied carbon for their upstream customers.  Further research is needed to accurately account for and verify embodied carbon across complex supply chains, along with regulations to expose such data in a standard format.

Carbon accounting frameworks are necessary for monitoring each company's carbon footprint, and incentivizing them to reduce it through social pressure. However, current frameworks do not ensure the exclusive attribution of carbon to a single entity, i.e., that every gram of carbon emitted is attributable to one and only one company, which leads to the scale mismatch between embodied and operational emissions.    One possible framework such exclusive attribution would be to make each company responsible for some percentage of their operational carbon, and make upstream companies in their supply chain, which purchase their products/services, responsible for the rest.  In this case, only a fraction of each company's operational carbon would be passed up the supply chain.  By adjusting the percentage of operational carbon passed upstream as embodied carbon, this framework could control the incentive for reducing operational versus embodied carbon.  Recently others have recognized this problem, and proposed similar frameworks that ensure the exclusive attribution of carbon emissions~\cite{article7}.  Such a framework would eliminate pitfalls \#2, \#3, and \#4, which ultimately derive from counting carbon emissions multiple times.  However, it may complicate pitfall \#1, since such a framework would also be difficult to monitor and enforce.   

\section{Conclusion}

This paper discusses general issues with optimizing for embodied carbon that are applicable to computer systems but broadly apply to the entire economy. Given the lack of direct financial incentive to reduce carbon emissions, it is important that the computing research community understand the implications of carbon accounting frameworks on carbon optimizations.  As we discuss, existing carbon accounting frameworks are complex and can lead to numerous pitfalls that may lead the research community down the wrong path. These issues are important because the community's view of carbon metrics will ultimately shape the field. 

\section*{Acknowledgements}
We thank the HotCarbon reviewers for their valuable comments, which improved the quality of this paper. This research is supported by NSF grants 2213636, 2136199, 2106299, 2102963, 2105494, 2021693, 2020888, 2045641, as well as VMware.

\balance
\bibliographystyle{ACM-Reference-Format}

\end{document}